\documentstyle[12pt]{article}
\begin{document}

\title{Thermodynamics of black holes in finite boxes}

\author{P.S.Cust\'odio and J.E.Horvath\\
\it Instituto de Astronomia, Geof\'\i sica e Ciencias Atmosf\'ericas\\
Rua do Mat\~ao, 1226, 05508-900 S\~ao Paulo SP, Brazil\\
Email: foton@astro.iag.usp.br}

\begin{abstract} We analyze the thermodynamical behavior of black
holes in closed finite boxes. First the black hole mass evolution
is analyzed in an initially empty box. Using the conservation of
the energy and the Hawking evaporation flux, we deduce a minimal
volume above which one black hole can loss all of its mass to the
box, a result which agrees with the previous analysis made by
Page. We then obtain analogous results using a box initially
containing radiation, allowed to be absorbed by the black hole.
The equilibrium times and masses are evaluated and their behavior
discussed to highlight some interesting features arising. These
results are generalized to $N$ black holes + thermal radiation.
Using physically simple arguments, we prove that these black holes
achieve the same equilibrium masses (even that the initial masses
were different). The entropy of the system is used to obtain the
dependence of the equilibrium mass on the box volume, number of
black holes and the initial radiation. The equilibrium mass is
shown to be proportional to a {\it positive} power law of the
effective volume (contrary to naive expectations), a result
explained in terms of the detailed features of the system. The
effect of the reflection of the radiation on the box walls which
comes back into the black hole is explicitly considered. All these
results (some of them counter-intuitive) may be useful to
formulate alternative problems in thermodynamic courses for
graduate and advanced undergraduate students. A handful of them
are suggested in the Appendix.
\end{abstract}

\maketitle

\section{Introduction}

In a seminal work of the 70\'s, Hawking \cite{haw}, showed that
black holes are capable of emitting radiation, and evaluated the
quantum process by which these black holes lose mass due to vacuum
polarization (induced by the gravitational energy of the black
hole). This important work and follow-up contributions, together
with Bekenstein's arguments \cite{bek} about the entropy of black
holes, prompted a new discipline of black hole research:
thermodynamics of black holes. Today, the study of black hole
thermodynamics is very active and several arguments indicate that
it must embrace the partial (or perhaps complete) marriage of
quantum mechanics, thermodynamics and gravitation.

The original Hawking analysis showed that the particles created
display a thermal spectrum $ f(E,T)={1\over{e^{\beta{E}}\pm{1}}}$
with $\beta=1/k_{B}T$, and a temperature determined by the black
hole mass $M$ given by

\begin{equation}
T_{bh}={{\hbar}c^{3}\over{8\pi{k_{B}}GM}} \sim 10^{-7} {\biggl(
{M_{\odot} \over{M}} \biggr)} \, K \label{1}
\end{equation}

see \cite{FN} for details. Since this spectrum is thermal, the
luminosity of the black hole can be calculated using the
Stefan-Boltzmann law, using  $L\propto{T^{4}{r_{g}}}^{2}$, where the
gravitational radius $r_{g}\propto{M}$ squared plays the role of the emitting
area. Using these definitions, $L(M)\propto{M}^{-2}$ is readily obtained.

In addition to "purely academic" cases (e.g. black holes inside ideal boxes),
these emission (absorption) relations are in principle useful for investigating
the properties of astrophysical and/or primordial black holes (hearafter PBHs) and
their cosmological consequences. For example, some short-timescale
gamma-ray burst events have been possibly explained by evaporating
PBHs in a terminal (explosive) phase. In fact, the derived relations between period,
mass and spectra are consistent with the PBH model, see \cite{grb}.

Even in the simplest cases  (the black holes interacting with ambient
radiation) it is interesting to understand what determines the
equilibrium features of the system and how is it achieved.
We shall perform an analysis restricted to finite volumes with the aim of
illustrating some novel
features of the thermodynamics of black holes using
{\it gedanken} experiments. These excercises serve also as
a prelude to the more complicated
situation, in which evaporation of PBHs must be studied in an expanding
universe. However, and quite independently of these further advanced
considerations, "boxed" black holes display interesting features
which need careful clarification.

\section{ Classical absorption in an initially cold box}

Let us consider the following {\it gedanken} experiment: we start
with a black hole and put it within a finite box (with linear size
$L$). In the beginning, the box has no radiation, i.e. its
temperature is zero. Then, since the black hole is hotter than the
environment, it will emit thermal radiation, and thus this closed
box will be increasingly filled with this radiation shortly
afterwards.

The evolution of the mass of the black hole is described by the well-known
expression which balances the energy loses to the Hawking luminosity
$\propto M^{-2}$

\begin{equation}
{dM\over{dt}} = - {C\over{M^{2}}} \label{2}
\end{equation}

with an initial condition of the box $T_{rad}(t=t_{i})=0$. The
constant $C$ depends on the degrees of freedom allowed to be
radiated by the black hole \cite{haw} and is set to $C
\sim{10}^{26}g^{3}{s}^{-1}$ throughout this work, and the initial
mass is $M_{i}$. We assume that the box linear size is much larger
than the black hole gravitational radius $L \gg r_{g}$ to avoid a
complicated non-linear feedback between the geometry and the
radiation field, which would lead us far beyond the purely
thermodynamical approach.

Since the box is closed, the total energy is conserved. Therefore
it is easy determine a relation for the black hole mass and the
thermal radiation contained within the box from

\begin{equation}
\varrho_{rad}(T_{rad})V_{box}+M(t) = M_{i} \label{3}
\end{equation}

where $t>t_{i}$ (we use natural units to set the speed of light
$c$ equal to 1). The radiation density is given by
$\varrho_{rad}(T)=a_{*}T^{4}$ with
$a_{*}\sim{8\times{10}^{-36}g{cm}^{-3}{K}^{-4}}$. After some time,
the box temperature is

\begin{equation}
T_{rad}(t)={\biggl[{M_{i}-M(t)\over{a_{*}V_{box}}}\biggr]}^{1/4}
\label{4}
\end{equation}

At an even later time $t > t_{i}$, eq.(\ref{2}) should not simply
contain the Hawking term, but an absorption term is also needed.
The simplest form of term is generally constructed as the product
of the incoming flux times the gravitational cross-section of the
radiation falling into the hole, $B\varrho_{rad}(T)M^{2}$, with
$B={27\pi{G}^{2}\over{c^{3}}} (or {27\pi\over{{M_{pl}}^{4}}}$ in
natural units, see Ref. \cite{CH} for details and discussion).
Therefore the complete differential equation for the mass is given
by

\begin{equation}
{dM\over{dt}}=-{C\over{M^{2}}}+B\varrho_{rad}(T)M^{2} \label{5}
\end{equation}

From this equation, it is clear that thermodynamical equilibrium
between the black hole plus and the ambient radiation will be
given by $\dot{M}=0$, and thus the mass of the black hole in
equilibrium is

\begin{equation}
M_{bh}=M_{c}(t_{eq})= {D\over{T_{rad}(t_{eq})}}
\end{equation}

where $D={(C/a_{*}B)}^{1/4} \sim{2\times{10}^{26}gK}$. Solving the
set of equations above we track the black hole mass evolution as
it approaches the equilibrium (and determine the conditions for
this equilibrium).

If we substitute the eq.(\ref{4}) into eq.(\ref{5}) we obtain

\begin{equation}
{dM\over{dt}}=-{C\over{M^{2}}}+B{\biggl({M_{i}-M\over{V_{box}}}\biggr)}M^{2}
\label{7}
\end{equation}

The evolution of the mass of eq.(\ref{7}) will be given by
solving the integral

\begin{equation}
\int_{M_{i}}^{M(t)}{dM{M}^{2}\over{-\kappa_{1}+\kappa_{2}M^{4}+\kappa_{3}M^{5}}}
= \, t \label{8}
\end{equation}

and the constants $\kappa_{i}$ are $\kappa_{1}=C$,
$\kappa_{2}=BM_{i}/V_{box}$ and $\kappa_{3}=-B/V_{box}$.

In the limit $V_{box}>>{r_{g}}^{3}$ the solution of eq.(\ref{7})
approaches $M(t)\sim{M_{i}{[{1-(t/t_{evap})}]}^{1/3}}$ with
$t_{evap}={M^{3}_{i}\over{3C}}$, as expected.

The numerical solution of the integral eq.(7) is shown in Fig.1
for different box sizes but the same initial black hole mass
$M_{i}$. In order to obtain the conditions for the equilibrium we
solve the algebraic equation $T_{rad}(t_{eq})=T_{Haw}(M_{eq})$,
rewritten as

\begin{equation}
{\biggl[{M_{i}-M(t_{eq})\over{a_{*}V_{box}}}\biggr]}^{1/4}={C\over{M(t_{eq})}}
\label{9}
\end{equation}

Inserting the respective constants above, eq.(\ref{9}) becomes

\begin{equation}
[1-\mu(t_{eq})]{\mu^{4}(t_{eq})} = D (V_{box}/{cm}^{3}) \times
{(M_{\odot}/M_{i})}^{5} \label{10}
\end{equation}

where $\mu(t_{eq})={M(t_{eq})\over{M_{i}}}$ and
$D\sim{4.2\times{10}^{-97}}$.

The interpretation of the solution is quite evident, when the
volume of the box is such that the right side of eq.(\ref{9}) is
larger than one, the black hole evaporates completely in the
absence of additional quantum corrections at the Planck scale. But
if $D (V_{box}/{cm}^{3}) \times (M_{\odot}/M_{i})^{5} \, \sim \,
1$, the size of the box is small enough to {\it stop} the
evaporation, and the system black hole$+$radiation achieves
thermodynamical equilibrium before its mass vanishes completely.

The linear size of the box below which the black hole achieves its
equilibrium is given by

\begin{equation}
L_{c}(M_{i}) \sim {1.3\times{10}^{32}} {(M_{i}/M_{\odot})}^{5/3}
\, cm \label{11}
\end{equation}

The time for reaching equilibrium $t_{eq}$ is determined by the
initial mass and the size of the box only, i.e.
$t_{eq}=t_{eq}(M_{i},V_{box})$. For the same initial mass, smaller
boxes will contain  black holes with larger final masses at
equilibrium (see Fig. 1). Our analysis agrees with the previous
work by Page \cite{pag}. From now on, we define the {\it critical
volume} by $V_{crit}(M)={L_{c}(M)}^{3}$ using eq.(\ref{10}), with
its clear physical interpretation given above. In the next
sections we revisit the derivation of eq.(\ref{10}), extend it to
more complicated cases and consider causality features.

\begin{figure}
      \caption{The approach to equilibrium of black holes in
finite boxes. The curves represent qualitatively the temporal
behavior of the mass of the black hole for two different values of
the box size $V_{box}$ (initially devoid of radiation); with the
initial value of the mass $M_{i}$ held fixed.}
         \label{Fig1}
   \end{figure}

\section{ Classical absorption in a closed box with initial radiation}

Let us now evaluate the behavior of the black hole mass when
introduced in a box with a non-zero initial radiation content. We
expect the black hole to achieve thermodynamical equilibrium
earlier than in the case without initial radiation, and with a
smaller equilibrium mass (considering boxes with the same size).
In this case, is easy show that

\begin{equation}
\int_{M_{i}}^{M(t)}{dM{M}^{2}\over{-\kappa_{1}+\kappa_{4}M^{4}(E_{i}-M)}}=t
\label{12}
\end{equation}

where $E_{i}=M_{i}+a_{*}{T_{i}}^{4}V_{box}$ and
$\kappa_{4}=B/V_{box}$. The plot of $M(t)$ as a function of time
is actually similar to the former case.

As before, we impose the equilibrium condition in to obtain the
relation between the black hole mass and the box volume. Following
the same steps as before, we obtain

\begin{equation}
{\mu^{*}(t_{eq})}^{4}[F(T_{i},M_{i})-\mu^{*}(t_{eq})]= D
{(V_{box}/{cm}^{3}) \times ({M_{\odot}/M_{i})}^{5}} \label{13}
\end{equation}

where $F(T_{i},M_{i})={E_{i}\over{M_{i}}}$, and the $t_{eq}$ is
different from the previous case. Actually the black hole achieves
thermodynamical equilibrium sooner, as expected.

\section{ $N$ black holes plus radiation in a closed box}

The generalization to $N$ black holes seems quite straightforward,
although it will become clear that the initial states and other
details must be carefully defined to achieve consistent results.
First, we shall consider just two black holes immersed in a closed
box (with constant volume) in the following initial situation: one
black hole has initial mass $M_{1}$ and the other black hole is
more massive than the first, $M_{2} > M_{1}$. Initially the box
has no radiation, and therefore these objects begin to evaporate
immediately. Moreover, we shall consider that
$V_{box}\sim{V_{crit}(M_{1})}$. Then at the initial time $t_{i}$
we have

\begin{equation}
{\biggl({dM_{2}\over{dt}}\biggr)}_{t_{i}}=-{C\over{M_{2}^{2}}}
\label{14}
\end{equation}

\begin{equation}
{\biggl({dM_{1}\over{dt}}\biggr)}_{t_{i}}=-{C\over{M_{1}^{2}}}
\label{15}
\end{equation}

Since $M_{1}<M_{2}$ one may consider $|(\dot{M_{1}})_{evap}| \gg
|(\dot{M_{2}}) _{evap}|$ initially. Afterwards, when the box
starts to be filled with the emitted radiation, some energy can be
absorbed by the black holes. Therefore, their evolution will be
given by

\begin{equation}
{{dM_{2}\over{dt}}}=-{C\over{M_{2}^{2}}}+B\varrho_{rad}(T){M_{2}}^{2}
\label{16}
\end{equation}

\begin{equation}
{{dM_{1}\over{dt}}}=-{C\over{M_{1}^{2}}}+B\varrho_{rad}(T){M_{1}}^{2}
\label{17}
\end{equation}

\begin{equation}
M_{1}(t)+M_{2}(t)+V_{box}{\varrho_{rad}(T)}=M_{1}(t_{i})+M_{2}(t_{i})
\label{18}
\end{equation}

where the last condition displays the conserved energy of the box.
Using this constraint, we can describe this system by the
following set of equations

\begin{equation}
{{dM_{2}\over{dt}}}=-{C\over{M_{2}(t)^{2}}}+{B\over{V_{box}}}{M_{2}(t)}^{2}
[E_{i}-M_{1}(t)-M_{2}(t) \label{19}
\end{equation}

\begin{equation}
{{dM_{1}\over{dt}}}=-{C\over{M_{1}(t)^{2}}}+{B\over{V_{box}}}{M_{1}(t)}^{2}
[E_{i}-M_{1}(t)-M_{2}(t)] \label{20}
\end{equation}

combining eqs.(\ref{17}) and (\ref{18}) and using the conservation
of energy yields

\begin{equation}
-C\biggl[{{1\over{M^{2}_{1}}}+{1\over{M^{2}_{2}}}}\biggr]+B\varrho_{rad}
[M^{2}_{1}+M^{2}_{2}]+V_{box}\dot{\varrho}_{rad}=0 \label{21}
\end{equation}

where $E_{i}=M_{1}+M_{2}$ is the initial energy.

Note that the set of coupled equations does not include the effect
of black hole motion due to their mutual gravity. Motion is likely
to affect the emission/absorption properties of the black holes in
the relativistic regime \cite{Ind}, \cite{PCJH}. Thus, the
analysis is strictly valid whenever $v_{bh} \ll c$.

Let us take a look at the global behavior of the solutions. First,
it is easy to show that the equilibrium between two black holes is
possible. If we impose that the thermodynamical equilibrium will
be achieved, then asymptotically

\begin{equation}
{{dM_{2}\over{dt}}}={{dM_{1}\over{dt}}}=0 \label{22}
\end{equation}

From these equations
${C\over{B}}V_{box}={M_{1}}^{4}[E_{i}-M_{1}-M_{2}]$ immediately
follows, and the same is true for the other black hole. Therefore,
it follows that

\begin{equation}
M_{1,eq}=M_{2,eq} \label{23}
\end{equation}

The same results of these equilibrium masses could have been
obtained from the expression of the entropy. In fact, more
complicated cases can be worked out either using the above
approach or using that the entropy must be an extreme for a system
in equilibrium. For instance, let  us consider $N$ black holes
enclosed in the box. The total entropy for this system is given by

\begin{equation}
S_{total}(t)={1\over{4}}\sum_{i}A_{i}(t)+g_{*}T^{3}_{rad}(t)V_{box}
\label{24}
\end{equation}

Since the horizon area for the i-th black hole is
$A_{i}=4\pi{r_{g, i}}^{2}$ we have

\begin{equation}
S_{total}(t)={4\pi\over{M_{pl}^{4}}}\sum_{i}M_{i}(t)^{2}+g_{*}T^{3}_{rad}(t)V_{box}
\label{25}
\end{equation}

In equilibrium
$T_{rad}=T_{bh1}=T_{bh2}=...=T_{bhi}={M_{pl}^{2}\over{8\pi{M_{eq}}}}$,
and therefore

\begin{equation}
S_{max}(N,V_{box})=\beta_{*}{N}{\mu_{eq}}^{2}
+{l_{*}/{\mu_{eq}}^{3}}(V_{box}/{cm}^{3}) \label{26}
\end{equation}

where $\beta_{*}=2.5\times{10}^{40}$,
$l_{*}=8\times{10}^{33}\alpha_{*}$ and $\mu_{eq}=(M_{eq}/10^{15}
\, g)$ the equilibrium mass scaled to a convenient reference
value.

Extremizing the entropy $\biggl({dS_{max}\over{dM_{eq}}}\biggr)=0$
and solving for $M_{eq}$ yields

\begin{equation}
M_{eq}(N,V_{box}) \sim 5.4 \times {10}^{13}
{\alpha_{*}}^{1/5}{N}^{-1/5}{(V_{box}/{cm}^{3})}^{1/5} \, g
\label{27}
\end{equation}

with $\alpha_{*}=26.7 g_{*}$. After substituting back into
$S_{max}$ we find

\begin{equation}
S_{max}(N,V_{box}) \sim 1.2 \times{10}^{38}
{\alpha_{*}}^{2/5}N^{3/5}(V_{box}/{cm}^{3})^{2/5} \label{28}
\end{equation}

Since $N=nV_{box}$ by definition, we have
$S_{max}\sim{1.2\times{10}^{38}}n^{3/5}(V_{box}/{cm}^{3})$.
Therefore, the thermodynamic approach leads us to conclude that
the equilibrium black hole mass is bigger for bigger boxes. This
result is quite counter-intuitive, but it has a reasonable
explanation given below. To confirm the dependence first, we
resort to another method.

Let us rewrite the equation for the mass of the i-th black hole as

\begin{equation}
{{dM_{i}\over{dt}}}=-{C\over{M_{i}(t)^{2}}}+{B\over{V_{box}}}{M_{i}(t)}^{2}[E_{i}-NM_{i}(t)]
\label{29}
\end{equation}

where we prepare the system with $N$ black holes with the same
initial mass (this particular condition is not over-restrictive
and simplifies the algebra). From  $\dot{M_{i}}=0$ we obtain

\begin{equation}
{\mu_{eq}}^{4}-{1\over{\mu_{i}}}{\mu_{eq}}^{5}=
{K\over{N\mu_{i}}}(V_{box}/{cm}^{3}) \label{30}
\end{equation}

where  $K \sim {3.5\times{10}^{-6}}$. This algebraic equation has
non-trivial solutions which can be found numerically. For the sake
of definiteness we adopt $N=1$, $\mu_{i}=10$, and a small
$V_{box}={10}^{3}{cm}^{3}$, which yields $\mu_{eq}\sim{0.137}$. If
we enlarge the box, say $V_{box}={10}^{4}{cm}^{3}$ instead, then
$\mu_{eq}\sim{0.244}$, and so on. If we fix the size of the box
instead, say $V_{box}=2.8\times{10}^{5}{cm}^{3}$, and change the
number of black holes, the equilibrium mass will reflect the
effect of the presence of other black holes accordingly. For
example, taking $\mu_{i}=10$ and $N=1$ we have
$\mu_{eq}\sim{1.027}$ but for $N={10}^{3}$ we obtain
$\mu_{eq}\sim{0.178}$ (Fig. 2). These results agree with the
previous entropy analysis, and confirm that the coupled equations
correctly describe the evolution of the system. It is quite
interesting to propose the construction of a plot showing the
behavior of the solutions and a discussion about the reasons for
that behavior.

\begin{figure}
 \caption{Graphical solution of the non-linear equation for the equilibrium
mass (eq.30) as a function of time for a fixed value of $V_{box}$
and a different number of black holes $N$. It may be said that the
equilibrium mass "feels" the presence of other black holes.}
\label{f3}
\end{figure}

As a further related work it is proposed to show that if we
consider the existence of an initial thermal radiation with energy
$E_{i}$, the effective volume becomes

\begin{equation}
V_{eff}(N,M_{i})={V_{box}\over{N+E_{i}/M_{i}}} \label{31}
\end{equation}

which holds for $N$ black holes with the same initial masses. This
effective volume must be used instead of $(V_{box}/N)$ in
eq.(\ref{26}) and $M_{eq}(N,V_{box})$ becomes
$\propto{[{V_{box}\over{N+E_{i}/M_{i}}}]}^{1/5}$.

The puzzling growth of the equilibrium mass in these systems must
be further explained in more detail to improve the understanding.
To proceed, let us consider two boxes with volumes $V_{box, 2}$
and $V_{box, 1}$ where the second box is larger than the first.
Let us immerse two black holes with the same initial masses in
both boxes. Therefore, the energy content is forced to be the
same. Then, in equilibrium we will have (dropping some inessential
numerical factors)

\begin{equation}
E_{box}=2 M_{eq, 1}+T^{4}_{rad, 1}V_{box, 1} = 2 M_{eq,
2}+T^{4}_{rad, 2}V_{box, 2} \label{32}
\end{equation}

Since the equilibrium mass is given by $M_{eq}
\propto{[V_{box}]}^{1/5}$ for both boxes, then we have

\begin{equation}
{V_{box, 1}}^{1/5} - {V_{box, 2}}^{1/5}=T_{rad, 2}^{4}V_{box, 2}
-T_{rad, 1}^{4}V_{box, 1} \label{33}
\end{equation}

both terms are negative due to the assumption $V_{box, 2}>V_{box,
1}$ then we obtain from the r.h.s. of this equation

\begin{equation}
{\biggl({T_{rad, 2}\over{T_{rad, 1}}}\biggr)}^{4}< \biggl({V_{box,
1}\over{V_{box, 2}}}\biggr)<1 \label{34}
\end{equation}

then $T_{rad, 2}<T_{rad, 1}$. Therefore, the bigger box will have
smaller radiation temperature. Since in equilibrium $T_{rad,
2}=T_{bh, 2} \propto {1\over{M_{eq, 2}}}$ then the mass of the
second black hole must be bigger than the first, in agreement with
our previous analysis. Previous work by Page \cite{pag}
(specifically his eq.(10)) has explored this problem, which has
been solved here using the conservation of energy only, with the
absorption terms explicitly present.

\section{Causality considerations}

The above considerations implicitly assumed that the black hole
absorption starts instantaneously as the former is created inside
the box. One might wonder whether a black hole immersed within an
initially empty box is actually able to absorb ambient radiation
immediately; since the Hawking radiation emitted by it will be
available to absorption only after $\sim L/c$ as required by
causality.

In order to gain some insight, we use the fact that if the box is
large enough, when the radiation comes back into the black hole,
the object has evaporated completely. Then, a critical condition
is obtained imposing this time interval to be bigger than the
timescale for evaporation. Thus

\begin{equation}
{L\over{c}} > {M^{3}\over{3C}} \label{35}
\end{equation}

Therefore, it follows that $L_{crit}(M)\sim
{{10}^{29}cm}{\biggl({M\over{10^{15}g}}\biggr)}^{3}$.

Then, any box (without initial radiation) containing a black hole
with mass $M$ whose linear size $L$ is bigger than the critical
size $L_{crit}$ above would let the black hole to evaporate
completely, thus precluding equilibrium. Note that the functional
dependence and numerical value are very different from the naive
approach used previously where we have considered that the
absorption term turns on instantaneously (that is, as soon we
immersed the black hole in the box). If we boldly compare
$L_{crit}$ with the size of the Hubble horizon today
$\sim{10}^{27}cm$, we deduce that a PBHs with masses smaller than
$\sim 2 \times {10}^{14}g$ can evaporate completely. However, it
is clear that in a realistic case other sources of energy must be
considered and our comment is intended just for pedagogical
purposes.

The result of eq.(\ref{35}) can be interpreted as follows. If we
put one hot black hole (at $t_{i}$) with initial mass $M_{i}$
inside a closed box with volume $V_{box}$, it will evolve
according to $\dot{M}=-C/M^{2}$ for a time $t_{i} < t <
{L_{box}\over{c}}$. But from $t > L_{box}/c$ on, the black hole
has to absorb radiation from the box (originated by itself!). Then
the absorption term $\propto{T^{4}V_{box}}$ contributes to avoid
its complete evaporation. If the box is slightly larger (but
smaller than the critical volume defined by eq.(10)) the onset of
absorption is delayed, then this black hole will be hotter than
the previous case. This black hole is then filling the box with
higher temperature radiation, and after some time it will be
absorbing radiation at higher temperature, hence the absorption
term $\propto{T^{4}M^{2}}$ will be larger. This term drives the
black hole into the larger equilibrium mass, as given by
eq.(\ref{27}). It should be remembered that a maximal number for
black holes enclosed in a box with volume $V_{box}$ must exist, a
naive estimate of $N_{max}$ can be obtained from the close-packing
condition $N_{max}r^{3}_{g}(M_{eq}) \sim {V_{box}}$.

\section{Conclusions}

In this work we have analyzed the behavior of black
holes+radiation systems, considering some {\it gedanken}
experiments within finite boxes (initially with and without
radiation) (see \cite{SWH} for a clear pioneering discussion of
these issues). These {\it gedanken} experiments are important to
understand the evolution of black holes in more complex
situations, since it is hoped that we can always approximate the
thermodynamical behavior of the universe with that of a finite
box. In these simple cases the energy available for absorption by
the black holes is finite at all times and there is no need to
consider other sources. Strictly speaking, and with an eye on the
actual cosmological formation and evolution of black holes, it may
be argued that the problem is oversimplified, since actual
primordial black hole masses must be formed with a substantial
fraction of the horizon mass \cite{carr}, and therefore finite
size effects should be considered from the scratch. The actual
situation is much more complicated but not impossible to tackle.
The real difficulty, which can be viewed as an advanced exercise
dealing with the (generalized) second law of thermodynamics not
related to laboratory systems, is that (unlike laboratory gases)
the black hole "gas" will have varying masses and therefore a
complicated kinetic equation must be used to describe their
evolution. This provides a concrete example, possibly realized in
nature, of a variable-mass gas needing a deep understanding of
statistical mechanics principles for its very formulation. The
simplest approach taken here was enough to show that one black
hole plus radiation can achieve thermodynamical equilibrium if the
box volume is smaller than a critical volume (in agreement with a
previous treatment made by Page \cite{pag}). Some surprises arise
in more general cases where we have two (or more) black holes
(plus radiation) within these boxes.

Even in their simplest versions the cases of black holes in boxes
are very instructive to analyze and may serve as an good issues
for a graduate course of thermodynamics/statistical mechanics (see
\cite{PMc} for a discussion of related formal aspects of black
hole thermodynamics) as an alternative to traditional problems
dealing with the approach to equilibrium. They pose several
questions of deep physical meaning, which are also strongly
entangled and sometimes puzzling to interpret.

\section{Appendix}

A complementary set of simple problems intended to reinforce the
conceptual aspects of black hole evaporation. Except for the
problem 4, their mathematical complexity is quite straightforward.

\bigskip
1) Using eq.(2), prove that the timescale for evaporation for one black hole is proportional to
$M^{3}$. Verify that for one black hole with the solar mass ($M \sim {2\times{10}^{33}g}$);
$t_{evap}(M) \sim{10}^{65}years$. Evaluate this for a initial mass $\sim{10}^{15}g$ and compare
with the age of the universe.

2) Verify the expression eq.(6) for the critical mass. Evaluate this number today,
considering that $T_{rad}(t_{0})\sim{3K}$.

3) Derive eq.(11). Use for the linear size of the box the present size of the universe.
Evaluate the corresponding mass at equilibrium.

4) Deduce the set of eqs.(19)-(21) and interpret the solutions
graphically.

5) Deduce the $M_{eq}(N,V_{box})$ for the cases without initial
radiation and with a radiation filling the box. Interpret these
results and compare with the cosmological situation plugging
representative numbers for the latter.

6) Discuss how all these results may be modified if the box expands following a known temporal dependence.

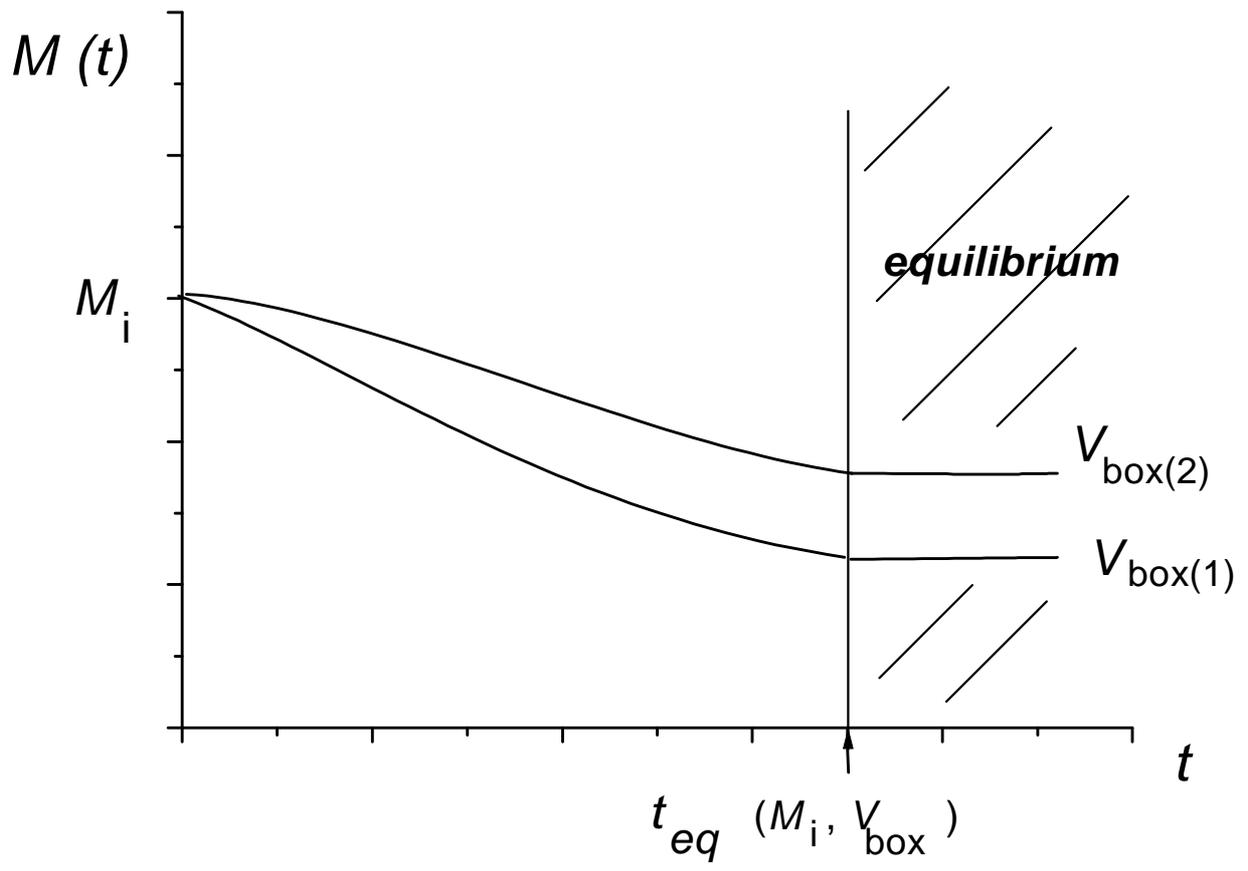
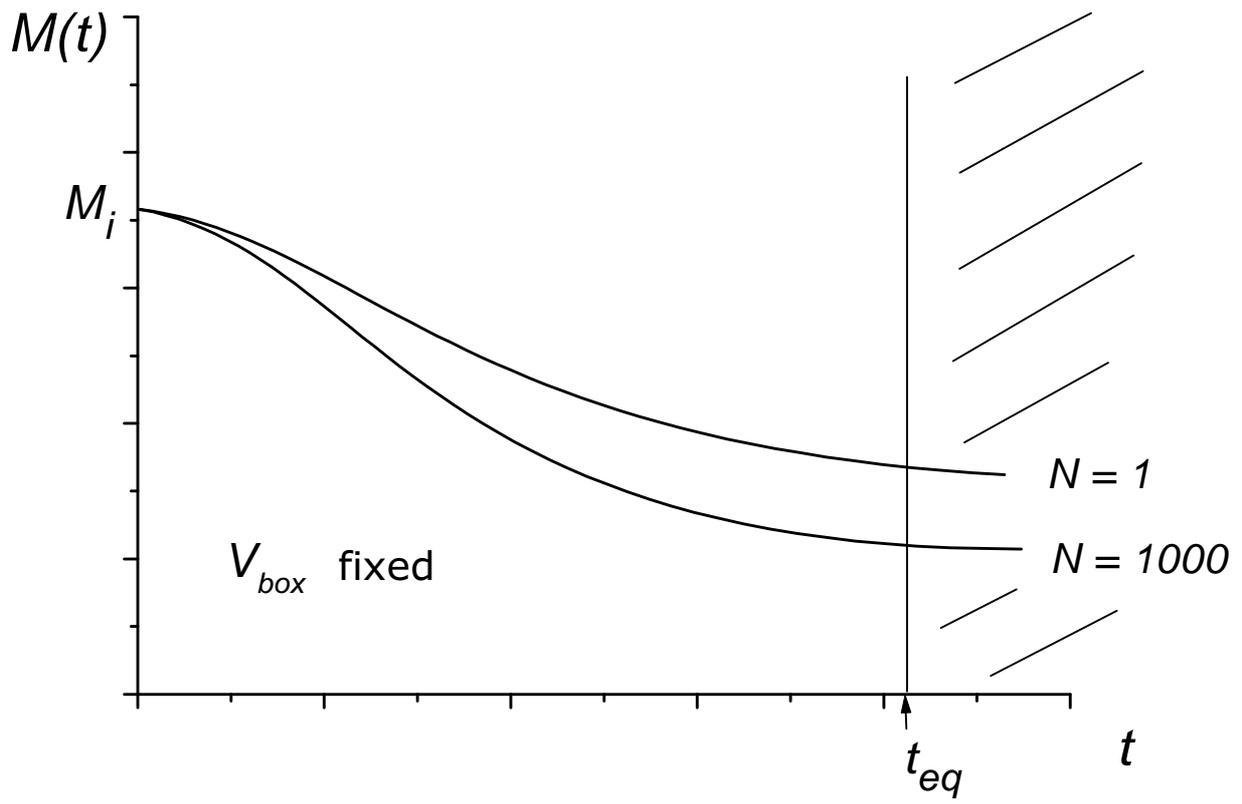
\end{document}